%% file: sample-authordraft.tex
\documentclass[a4paper]{article}

\usepackage[english]{babel}
\usepackage[utf8x]{inputenc}
\usepackage[T1]{fontenc}

\usepackage[a4paper,top=3cm,bottom=6cm,left=3cm,right=3cm,marginparwidth=1.75cm]{geometry}

\usepackage[table]{xcolor}
\usepackage{booktabs}
\usepackage{amsmath}
\usepackage{graphicx}
\usepackage[colorinlistoftodos]{todonotes}
\usepackage[colorlinks=true, allcolors=blue]{hyperref}
\usepackage{authblk}
\usepackage[comma,super,sort&compress]{natbib}
\usepackage{nomencl}
\usepackage{booktabs}
\usepackage{tabularx}
\usepackage{textcomp}
\usepackage{amsfonts}
\usepackage{subfig}
\usepackage[version=4]{mhchem}
\usepackage{siunitx}
\usepackage{xcolor}
\date{}

\usepackage{lineno}
\usepackage{color}
\usepackage{soul}
\usepackage{cancel}

\begin{document}

\title{Graph neural networks for materials science and chemistry}

\author[1,2]{Chen Shao}
\author[1]{Chen Zhou}
\author[1,3,*]{Pascal Friederich}

\affil[1]{Institute of Theoretical Informatics, Karlsruhe Institute of Technology, Am Fasanengarten 5, 76131 Karlsruhe, Germany}
\affil[2]{Present address: Institute for Applied Informatics and Formal Description Systems, Karlsruhe Institute of Technology, Kaiserstr. 89, 76133 Karlsruhe Germany}
\affil[3]{Institute of Nanotechnology, Karlsruhe Institute of Technology, Hermann-von-Helmholtz-Platz 1, 76344 Eggenstein-Leopoldshafen, Germany}
\affil[*]{Contact: pascal.friederich@kit.edu}

\title{Graph neural networks to learn joint representations of disjoint molecular graphs}


\maketitle

\begin{abstract}
Graph neural networks are widely used to learn global representations of graphs, which are then used for regression or classification tasks. Typically, the graphs in such data sets are connected, i.e. each training sample consists of a single internally connected graph associated with a global label. However, there is a wide variety of yet unconsidered but application-relevant tasks, where labels are
assigned to sets of disjoint graphs, which requires the generation of global representations of disjoint graphs. In this paper, we present a new data set with chemical reactions, which is illustrating this task. Each sample consists of a pair of disjoint molecular graphs and a joint label representing a scalar measure associated with the chemical reaction of the molecules. We show the initial results of graph neural networks that are able to solve the task within a combinatorial subset of the data set, but do not generalize well to the full data set and unseen (sub)graphs.
\end{abstract}

\section{Introduction}\label{sec:intro}
\input{files/intro}

\section{Related Works}
\input{files/related_work}

\section{Dataset of Reaction}
\label{sec:dataset of reaction}
\input{files/dataset}

\section{Experimental Setup}
\label{sec:exp setup}
\input{files/experimental_setup}

\section{Results}
\label{sec:result}
\input{files/result}

\section{Conclusions}
\input{files/conclusion}

\section*{Acknowledgement}
The authors acknowledge support by the state of Baden-Württemberg through bwHPC.

\bibliographystyle{naturemag}
\bibliography{sample-base}

\appendix 
\input{files/appendix}

\end{document}

%% file: files/intro.tex
In the last few years, graph neural networks (GNNs) attracted growing attention in chemical sciences, where they play an important role in solving challenges in molecular property prediction and design.
Currently, GNNs are widely used for regression and classification tasks, e.g. to predict molecular solubility or toxicity.
However, most currently considered tasks of GNNs are limited to single input molecular graphs, i.e. they use internally connected input molecular graphs to learn node and edge representations, convert them to global graph representations, from which the global label is then predicted.
However, there is a wide range of real-world tasks where global labels are assigned to sets of input graphs, rather than single input graphs.
Examples for such tasks is the prediction of solubility (not for a single solvent but for arbitrary combinations of solvents and solutes), reactivity prediction, where two or more molecular graphs react and the task is to identify the reaction center or a global property such as the reaction energy or reaction barrier, catalytic activity, e.g. of catalytic surfaces and given reactants, and many more.
\textbf{Common in all those tasks is that the joint label depends not on one but on multiple disjoint input graphs.}

\begin{figure}
	\centering
	\includegraphics[width=0.91\textwidth]{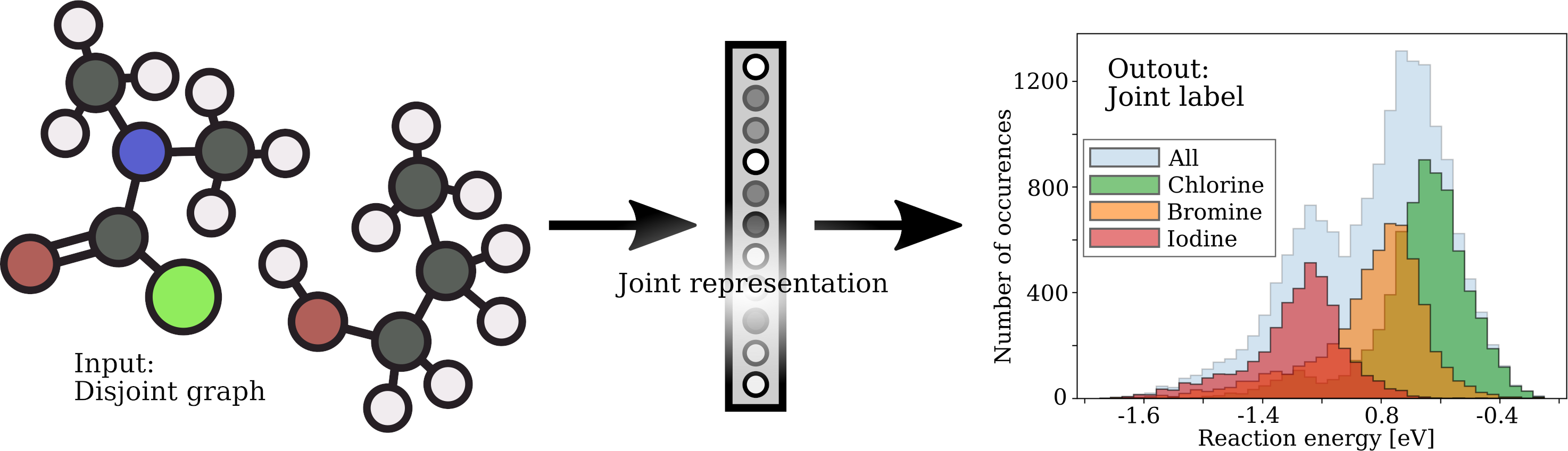}
	\caption{New task for graph learning: Learning joint representations to predict global labels from sets of disjoint input graphs.}
	\label{fig:overview1}
\end{figure}

Many GNNs for tasks in chemistry depend not only on the connectivity of the nodes in the input graph but also on their geometric arrangement. The labels in the task we are presenting here are typically invariant to the relative geometric arrangement of the different molecular graphs. However, the internal geometry of each of the input graphs might still add useful information about the final label, even though the geometry follows from the connectivity and the node features, and thus might as well be learned, given enough data. The development of GNN architectures for the task of learning global representations of disjoint graphs is therefore a highly relevant research area, and we hope that this work stimulates further development in that direction.

%% file: files/related_work.tex
The prediction of molecular properties is a cornerstone in chemistry, e.g. in drug discovery, where accurate predictions are needed to identify drug candidates in an efficient and computationally inexpensive way.
Molecular graphs allow us to learn informative representations of molecules, by learning from the chemical structure of a molecule directly and enhancing that information with physics-informed features of atoms and bonds, potentially including the 3D geometry information of the molecular structure.
The basic principle of GNNs is as follows: Atoms that are connected by bonds are close in the graph, which means that they have the greatest influence on each other. Through graph convolutions or message passing, such pairwise influence decays with the distance between the atoms. This enables GNNs to learn informative atom representations which can then be combined to global vector representations of entire molecules.

GNNs have found tremendous success in processing molecules and molecule properties, which has become one of their main applications \cite{zhou2021graph}. Seminal work by Duvenaud {\it et al.} showed how GNNs can be seen as a generalized and learnable alternative to until then prevalent fingerprint representations of molecules.\cite{Duvenaud2015} Gilmer {\it et al.} suggested a more generalized framework which they called message passing neural networks (MPNNs) and showed that MPNNs can accurately predict quantum mechanical properties, calculated by density functional theory (DFT), which allows the wider and more successful application of GNN to quantum chemistry.\cite{gilmer2017neural} Nowadays, many GNN architectures are available in the hope of being able to replace expensive quantum mechanical calculations with fast data-driven predictions.\cite{bouritsas2021improving} \cite{corso2020principal} \cite{ beaini2021directional} \cite{bodnar2021weisfeiler}.

Each molecule and associated molecular property can be uniquely determined by its 3D representation. After seminal and very promising work in that direction,\cite{gilmer2017neural}, Schütt {\it et al.} \cite{schuett2017schnet} leveraged continuous-filter convolutions to learn local atomic environments in an architecture consisting of atom-wise blocks and interaction blocks.
This idea is further optimized in DimeNet and DimeNet++ to achieve a new level of accuracy for quantum mechanical property predictions.\cite{klicpera2020directional}
Flam-Shepherd {\it et al.} \cite{flam2021neural} introduced higher-order paths to incorporate angle and dihedral information into a message passing framework, while Klicpera {\it et al.} introduced synthetic coordinates and directional message passing to leverage angular information.\cite{klicpera2021directional}
Spherical Message Passing (SMP) \cite{liu2021spherical} inherits this and further adds more angular information to completely encode all the information of its 1-hop neighbors. In a similar fashion, SMP was superseded by GemNet \cite{klicpera2021gemnet} that finally also captures torsion angles (the relative rotation around a bond of two substructures connected by that bond). With this inclusion, the whole geometry of a molecule is uniquely defined and the MPNN can leverage all of the 3D information.
In this line of work, SE(3)-equivariant GNNs \cite{batzner2021se3equivariant} have to be mentioned as well which uses the 3D structure in an equivariant architecture for simulating a molecule’s dynamics.

Overall, this summary shows the impressive amount and progress in research on 3D graph representation learning. While covering many important challenges in chemistry, there are important tasks that are not covered, including the prediction of properties which only indirectly dependent on the exact 3D structure of molecules, and - more importantly - which depend on more than one input graph.

One of the most prominent examples for learning from disjoint input graphs is the area of reaction prediction and retrosynthesis. Research in this area includes the graph transformation policy network for chemical reaction prediction \cite{do2019graph}, where graph neural networks are combined with reinforcement learning to identify graph modifications (on disjoint input graphs) which indicate potential reactions between input molecules. Wen {\it et al.} use graph neural networks to learn molecule embeddings of reactants and products which are then concatenated to predict bond dissociation energies, rather than learning joint representations of all relevant (disjoint) molecules.\cite{wen2021bondnet} Wang {\it et al.} train graph neural networks to learn molecular graph representations, in a way that the sum of molecular embeddings of reactants corresponds to the embedding of the product of the reaction.\cite{wang2021chemical} This method very elegantly avoids the need for joint embeddings but is also limited as reaction-specific properties such as reaction energies and conditions cannot easily be predicted.

%% file: files/dataset.tex
The data set introduced here is an illustration of a very common, yet not fully explored class of tasks of graph learning in chemistry, i.e. learning global labels of disjoint molecular graphs.
A wide range of other data sets illustrating the same class of tasks are relevant to applications and will be developed, however, they are not required here to illustrate the task.

The data set presented here was constructed in a combinatorial way.
We focus on the reaction between two families of chemicals, alcohols and acyl halides. We generated lists of molecules of each family and calculated the reaction energy difference using semi-empirical quantum mechanical calculations (using the xTB software \cite{bannwarth2021extended}) between pairs of molecules to construct our data sets. More information can be found in Appendix~A.

In principle, data sets with arbitrary numbers of disjoint input graphs are thinkable, but as a proof-of-principle, we focus on data points with only two input graphs here. In total, our full data set II contains 16599 data points, each of which is a combination of two molecules drawn from 299 different alcohols and 285 different acyl halides. Data set I is a subset of data set II, with a smaller number of alcohols and acyl halides. The labels are distributed according to a mixture of three slightly asymmetrical Gaussian distributions with similar standard deviations, but different means (one for reactions with bromine, chlorine, and iodine, respectively, see Fig.~\ref{fig:overview1}).

In all experiments, the energy difference is normalized using mean and standard deviation. We use the 3D coordinates of all atoms as edge features in graph neural networks to capture geometrical information.

%% file: files/experimental_setup.tex
\begin{figure}[t]
\centering
\begin{minipage}[t]{0.3\linewidth}
\centering
\caption*{(a)}
\includegraphics[width=0.7\textwidth]{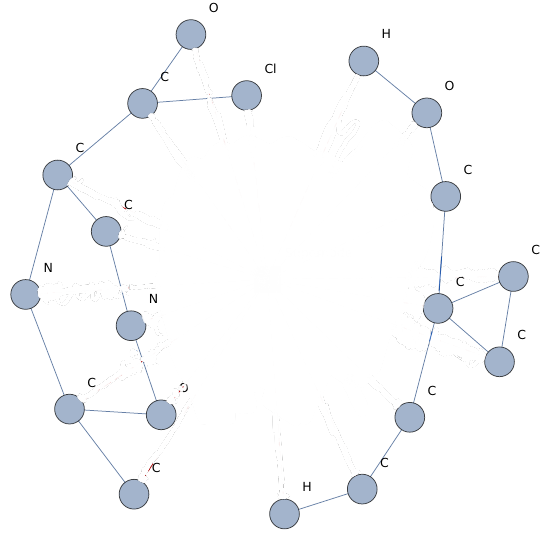}
\end{minipage}
\begin{minipage}[t]{0.3\linewidth}
\centering
\caption*{(b)}
\includegraphics[width=0.7\textwidth]{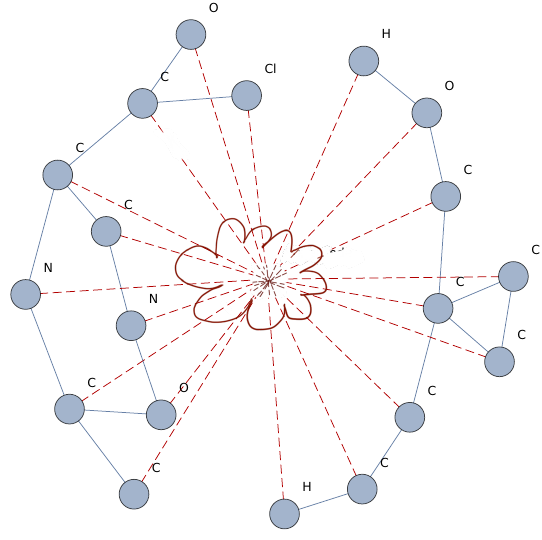}
\end{minipage}
\begin{minipage}[t]{0.3\linewidth}
\centering
\caption*{(c)}
\includegraphics[width=0.7\textwidth]{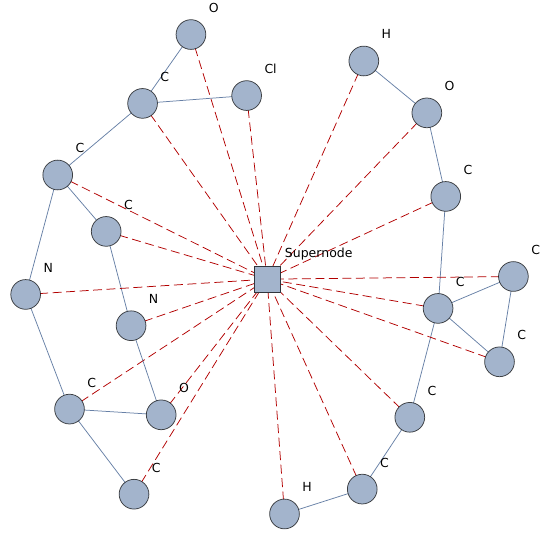} 
\end{minipage}
\centering
\caption{Visualization of three different input representations: (a) disjoint graph (DG), (b) fully connected graph (FC), and (c) global node (GN).}
\label{fig:visualiz ation method}
\end{figure}

In order to explore how existing graph neural networks perform on the data set presented in this work, we implemented the following experiments:
\begin{itemize}
    \item Experiment 1: On random splits of a combinatorial subset of the data (to test interpolation capabilities), we trained graph neural networks on a) disjoint graphs, where the graph embeddings are concatenated after the global pooling step ("disjoint graph", DG), b) on fully connected graphs, where the edge embeddings carry information whether edges are chemical bonds or additional non-physical connections ("fully connected graph", FC), and b) on disjoint input graphs with one additional global node to which all other nodes are connected to ("global node", GN) (see Fig.~\ref{fig:visualiz ation method}).
    \item Experiment 2: To test generalization to unseen input graphs, we trained models on splits of the data that ensure that a) each type of alcohol molecule is only present in either training, validation or test set. The same can also be done for acid halide molecules, which is not shown in this work. We ensured that all three types of halides (Cl, Br, I) are present in the training set to ensure that no new node types (chemical elements) occur during validation.
\end{itemize}

\begin{figure}
    \centering
    \includegraphics[width=0.5\linewidth]{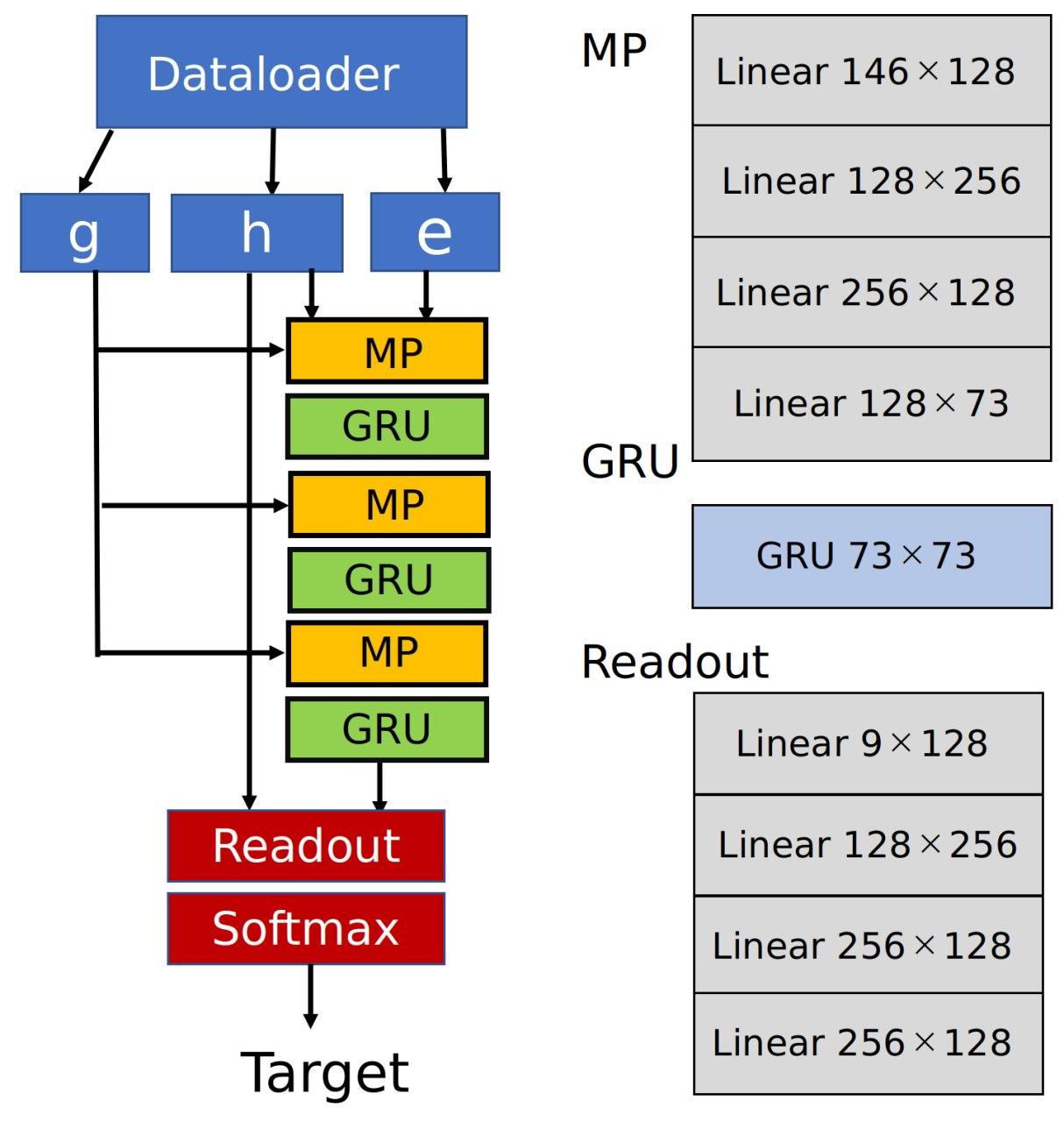}
    \caption{Illustration of the baseline message passing network in all experiments.\cite{gilmer2017neural}}
    \label{fig:architecture}
\end{figure}

The graph neural networks used in all experiments are based on the MPNN model proposed by Gilmer {\it el al.},\cite{gilmer2017neural} a illustration is given in Fig. \ref{fig:architecture} which has the mathematical formulation written as follows: 

\textbf{Message Passing Framework}: The message passing step can be formally written as:
\begin{equation}
    \mathbf{m}_v^{t+1}= \sum_{w\in \mathcal{N}(v)}A_t(\mathbf{e}_{vw})\mathbf{h}_w^t
    \label{equ: message passing}
\end{equation}
\begin{equation}
     \mathbf{h}_v^{t+1} = U_t(\mathbf{h}_v^{t}, \mathbf{m}_v^{t+1})
     \label{equ: updation}
\end{equation}
where $\mathbf{m}_v^{t+1}$ is the "message" aggregated from node $v$'s neighborhood $\mathcal{N}(v)$ at iteration $t$, $\mathbf{e}_{vw}$ is edge vector between node $v$ and $w$, and $\mathbf{h}_{w}^t$ is hidden feature vector of node $w$. 
$A_t$ and $U_t$ are arbitrary differentiable functions. $A_t$ is instantiated as three densely connected layers with activation function ReLU. $U_t$ is a single GRU Layer. 

\textbf{Pooling Layer}: The pooling layer for global graph embedding ("global readout") 
can be written as: 
\begin{equation}
    y = \sum_{v\in V}\sigma \left( i(\mathbf{h}_{v}^{T}, \mathbf{h}_v^0) \right) \cdot \left(j (\mathbf{h}_v^T)\right)
    \label{equ:readout}
\end{equation}
where $i$ and $j$ share the same architecture as $A(\mathbf{e}_{vw})$. $\mathbf{h}_v^0$ is the initial node embedding and $\mathbf{h}_v^{T}$ is the final node embedding after the last message passing step.

\textbf{Reaction-based pooling layer}: Based on the above-mentioned baseline, we have designed an adaptive pooling layer to improve the performance for reaction-level prediction. It can be formally written as: 

$\text{MPNN}_{\text{global}}$ follows the pooling function:
\begin{equation}
    y = \sigma \left( i(\mathbf{h}_g^T, \mathbf{h}_g^{0}) \right) \cdot \left(j (\mathbf{h}_g^T) \right)
    \label{equ:pooling global readout}
\end{equation}

where $g$ is the index of the global node $h_g$, this is abbreviated as global readout (GR).

\textbf{Concatenation pooling layer}: Furthermore, we implemented a more general concatenation-pooling layer, where a final representation for each node $v \in \mathcal{V}$ is computed using 
\begin{equation}
    \mathbf{h}_v^{\text{final}}= \textit{CONCAT}\hfill(\mathbf{h}_v^{(0)}, \mathbf{h}_v^{(1)}, \cdots, \mathbf{h}_v^{(T)})
    \label{equ: concat}
\end{equation}

The pooling layer can be written as:

\begin{equation}
    y = \sum_{v\in V}\sigma \left(i(\mathbf{h}_v^{\text{final}}) \right) \cdot \left(j (\mathbf{h}_v^T)\right)
    \label{equ:concat readout}
\end{equation}

\textbf{Normalization}: On top of adapting the global readout layer, we also modified the normalization in the following way: We normalize the initial node embedding row-wise and column-wise. The initial node embedding is based on physical node features, such as formal charge, which can have a much larger value than one-hot encoded node features. Utilization of the above-mentioned normalization method achieves minimized conditional number, which provides more numerical robustness in the training process.  \\

\textbf{Baseline MLP}:
As a baseline for comparison with the different MPNNs, an MLP was trained on concatenated fingerprint representations of the molecules.\cite{doi:10.1021/ci100050t} Morgan fingerprints as implemented in RDKit with a radius of 3 and a size of 1024 bits were used.

%% file: files/result.tex
\textbf{Experiment 1}: Comparing the performance of various graph neural networks with a basic MLP trained on concatenated molecular fingerprint representations, we find that the global node method is the most promising way of implementing graph neural networks to learn joint representations of disjoint graphs. As shown in Tab.~\ref{table:result1}, the global node (GN) and fully connected graph (FC) methods outperform the disjoint graph (DG) method. Moreover, the GNN  method achieved the best performance consistently on all metrics. The advantage of global node also empirically demonstrated the existing work 
showing that long-range interaction is crucial for GNN's generalization performance \cite{rampek2021hierarchical}. 
Because the global node's two-hop neighborhood contains all atoms, it directly improves the efficiency of modeling long-range interactions.

\begin{table}
\centering
\small
\begin{tabular}{ l | c c c c c c c}
     \toprule 
     \textbf{Method} & $\mathbf{r^2}$ & \textbf{RMSE} & \textbf{SRE} & \textbf{MAE}\\ 
     \hline 
     MPNN DG (Test) & 0.834 &  7.570e-1 & 8.840e-2 & 5.500e-2\\
      \hline 
      MPNN DG (Train) & 0.900 &  6.380e-2 & 5.670e-3 & 4.990e-2\\
      \hline 
      MPNN FC (Test) & 0.943 &  4.680e-2  & 3.600e-3 & 2.41e-2\\
      \hline 
      MPNN FC (Train) & 0.998 &  1.320e-2 & 3.600e-4 & 1.160e-2\\
     \hline
      MPNN GN (Test)  & \textbf{0.999} &  \textbf{4.200e-2} & \textbf{2.870e-3} & \textbf{2.020e-3}\\  
      \hline 
      MPNN GN (Train)  & 0.999 &  9.650e-3 & 2.020e-4 & 7.420e-3\\
      \hline 
      \hline 
     MLP (Test) & 0.856 & 9.124e-02 & 1.205e-02 & 6.779e-02 \\
     MLP (Train) & 0.869 & 8.740e-02 & 1.016e-02 & 6.409e-02 \\
     \bottomrule 
\end{tabular}
\caption{Experiment 1, performance of multiple different MPNNs compared to a MLP on data set I. The global node MPNN model (GN) outperforms the other methods disjoint graph (DG) and fully connected graph (FC) consistently, with a $r^2$ value which is $21.8\%$ larger than that of the MLP.}
\label{table:result1}
\end{table}

\textbf{Experiment 2}: In the second experiment, we have split the combined data set so that there are no overlaps in alcohol molecules between the training and test sets, i.e. all alcohol molecules in the test set were not seen during model training. Tab.~\ref{tab:conventional method} presents the performance of the MPNN with a global node and modifications thereof, compared again to MLP. It can be observed that a) the generalization task is much harder than the interpolation task in experiment 1, b) that the MLP outperforms the MPNN with global node, and c) that further modifications, i.e. improved normalization of node embeddings as well as a modified global readout method based on only the global node are necessary to obtain a better performance than the MLP.

\begin{table}
\centering
\begin{tabular}{ l | c c c c c c c}
     \toprule 
     \textbf{Method} & $\mathbf{r^2}$ & \textbf{RMSE} & \textbf{SRE} & \textbf{MAE}\\ 
     \hline 
      MPNN GN (Test) & 0.766 &  0.116  &  2.285e-2 & 8.926e-2\\
      \hline 
      MPNN GN (Train) & 1.000 & 2.895e-3 & 1.162e-5 & 2.256e-3\\
     \hline 
      +Norm (Test)  & 0.766 & 0.112  & 1.569e-2 & 8.523e-2\\
     \hline
      +Norm  (Train)  & 1.000 &  2.770e-3 & 1.121e-5 & 2.149e-3\\
      \hline 
      +Norm+CR (Test)  & 0.785 & 0.113 & 5.923e-2 & 8.267e-2\\  
     \hline
      +Norm+CR (Train)  & 1.000 &  3.183e-3  & 1.417e-5 &  2.460e-3\\
     \hline
      +Norm+GR (Test)  & \textbf{0.787} & \textbf{0.109}  & \textbf{1.492e-2} & \textbf{8.144e-2}\\  
     \hline
      +Norm+GR (Train)  & 1.000 &  3.609e-3 & 2.075e-5 & 2.793e-3\\
     \hline 
      MLP (Test) & 0.668 & 1.350e-01 & 6.795e-02 & 1.051e-01 \\
     MLP (Train) & 0.879 & 8.411e-02 & 1.373e-02 & 6.216e-02 \\
     \bottomrule 
\end{tabular}
\caption{Experiment 2, performance of multiple different MPNNs compared to a MLP on a generalization task based on data set II. We extended the MPNN GN model from Table~\ref{table:result1} with row-wise normalization (+Norm), and additionally tested alternative readout methods, i.e. readout from the global node embedding only (GR), as well as a more general version of the concatenation-readout (CR). The implementation of GR, CR and Norm are introduced in Section~\ref{sec:exp setup}.
}
\label{tab:conventional method}
\end{table}

%% file: files/conclusion.tex
We presented a combined data set of chemical reactions, focusing on the prediction of Gibbs free energy differences for a reaction of two molecules. Based on the data set, we defined a new sub-task for graph representation learning beyond the current limitation of a single graph: Learning global representations of disjoint graphs. We have demonstrated that message-passing neural networks, in particular with global nodes are able to learn such joint representations and predict global labels. However, their performance in a generalization scenario is only slightly better than the performance of a MLP. One reason might be that the hyperparameters of the MPNN are not optimally tuned for the task at hand (while the hyperparameters of the MLP are roughly optimized), but it might also be the case that the global node method is not optimal in the generalization task for disjoint graphs. Therefore, we illustrated that this data set is challenging for current graph neural network architectures, and we hope that our initial work encourages further research on the optimization of GNN architectures, particularly for disjoint graph representations, in order to extend their scope of application in chemistry. 

%% file: files/appendix.tex
\section*{Appendix} \label{sec:appendix}

\subsection*{Appendix A: Reaction data set details}\label{app:dataset}

The data set presented here was constructed in a combinatorial way.
We focus on the reaction of alcohols with acyl halides, and generated a list of alcohols, as well as three lists of acyl halides, with chlorine, bromine and iodine, respectively.
We split the alcohols in 6 subgroups, and each of the acyl halides in two subgroups respectively. Using combinations of those subgroups, we generated 6 combinatorial data sets.
Each data point consists of one alcohol $R_1-OH$ and one acyl halide molecule $R_2-X$ , where $R_1$ and $R_2$ are arbitrary chemical groups fulfilling certain side conditions (e.g. $R_1$ must not contain an acyl halide, and $R_2$ must not contain an $OH$ group), $OH$ is a hydroxyl group, and $X$ is a halogen atom (i.e. Cl, Br or I).
For each data point, we computed the reaction energy $\Delta E$, i.e. the energy difference between the reactants ($E(R_1-OH) + E(R_2-X)$) and the products ($E(R_1-O-R_2) + E(H-X)$) using semi-empirical tight-binding calculations using the GFN-xTB software.\cite{bannwarth2021extended}
This reaction energy $\Delta E$ serves as a label for the two input graphs $G(R_1-OH)$ and $G(R_2-X)$. A illustration of the data set 2 is given in Fig.~\ref{fig:distribution of dataset}.

\subsection*{Appendix B: Outlier detection}
As the data sets is constructed using semi-empirical quantum mechanical calculations, it is not free of errors, in this case failed calculations. Those typically lead to unreasonable high or low reaction energies which are easy to find and double-check with the initial calculations. In that way, we removed multiple obvious outliers from the data set.

\begin{figure}
    \centering
    \includegraphics[width=0.8\linewidth]{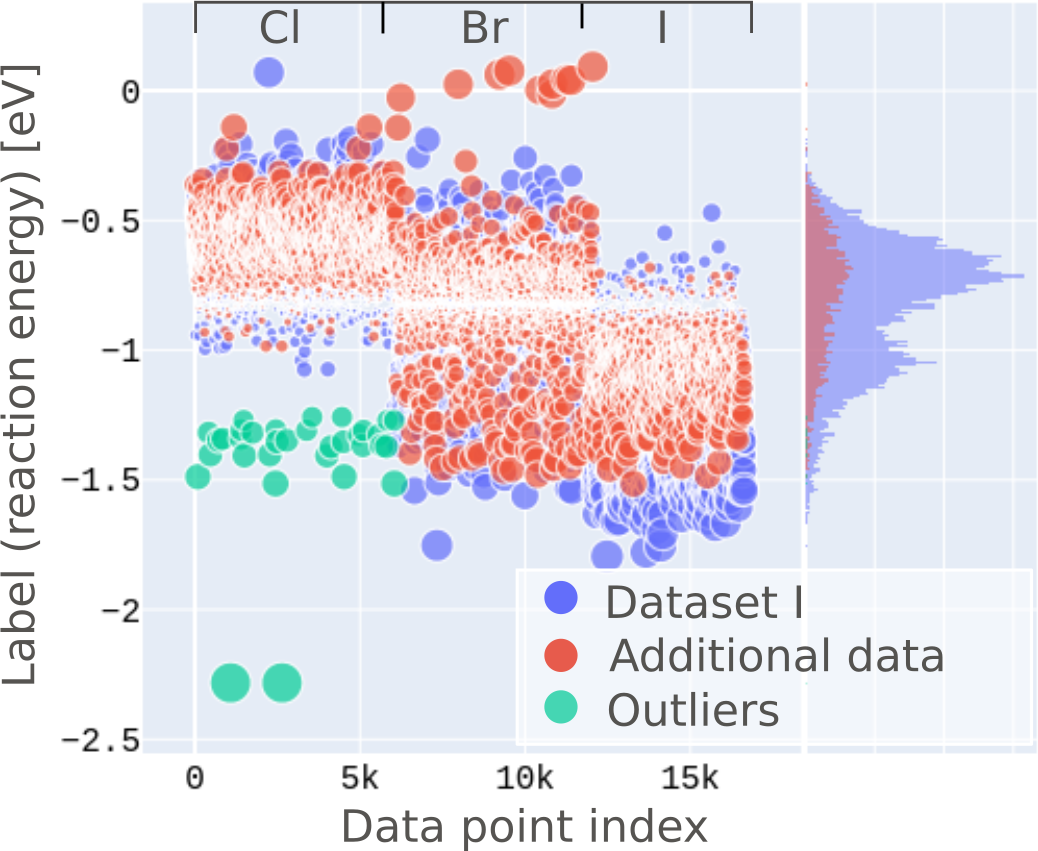}
    \caption{Distribution of energy differences in the two data sets, where data set II includes the data of data set I, with addition of additional molecules (both alcohols and acid halides).}
    \label{fig:distribution of dataset}
\end{figure}

However, as a test of the outlier detection capabilities, we also trained a MPNN model on a data set containing outliers. We detect the outliers through a scatterplot-based method, which means we evaluate the trained model on the new samples, and sort the data using the prediction error and define the samples possessing large errors as outliers. The outliers are then iteratively deleted while monitoring the training process. Figure~\ref{fig:distribution of dataset} shows a distribution of data points, where automatically found outliers are marked in green. Comparison with our manual outlier removal procedure shows that all outliers due to simulation failures were found.

